\documentclass{iaus}
\usepackage{ifpdf}
\ifpdf
  \usepackage[pdftex]{graphicx,color}
\else
  \usepackage[dvips]{graphicx,color}
\fi

\title[VLT-FLAMES Tarantula Survey] 
{The VLT-FLAMES Tarantula Survey}

\author[Evans et al.]   
{C.~J.~Evans$^1$, N.~Bastian, Y.~Beletsky, I.~Brott, M.~Cantiello, J.~S.~Clark, P.~A.~Crowther, A.~de~Koter, 
S.~de~Mink, P.~L.~Dufton, P.~Dunstall, M.~Gieles, G.~Gr\"{a}fener, V.~H\'{e}nault-Brunet, A.~Herrero, 
I.~D.~Howarth, N.~Langer, D.~J.~Lennon, J.~Ma\'{i}z~Apell\'{a}niz, N.~Markova, F.~Najarro, J.~Puls, H.~Sana, S.~Sim\'{o}n-D\'{i}az, 
S.~J.~Smartt, V.~E.~Stroud, W.~D.~Taylor, C.~Trundle, J.~Th.~van~Loon, J.~S.~Vink \and N.~R.~Walborn}

\affiliation{$^1$UK ATC, Royal Observatory Edinburgh, 
Blackford Hill, Edinburgh, EH9 3HJ, UK \\ email@ {\tt chris.evans@stfc.ac.uk} 
}

\pubyear{2009}
\volume{266}  
\pagerange{1-6}
\jname{Star Clusters}
\editors{Richard de Grijs \& Jacques R. D. L\'epine, eds.}
\begin{document}

\maketitle

\begin{abstract}
The Tarantula Survey is an ambitious ESO Large Programme that has
obtained multi-epoch spectroscopy of over 1,000 massive stars in the
30~Doradus region of the Large Magellanic Cloud.  Here we introduce
the scientific motivations of the survey and give an overview of the
observational sample.  Ultimately, quantitative analysis of every
star, paying particular attention to the effects of rotational mixing
and binarity, will be used to address fundamental questions in both
stellar and cluster evolution.

\end{abstract}

\firstsection 
\section{Introduction}

The Tarantula Nebula (30~Doradus, NGC\,2070) in the Large Magellanic
Cloud (LMC) is the brightest and most massive H~{\scriptsize II} region in the Local Group.
It is a beautiful and very intricate region, far removed from a
`simple single-stellar-population'.  Indeed, \cite{wb97} identified at
least five distinct populations \cite{w09}:

\begin{itemize}
\item{The central `Carina Phase' concentration, rich in early O-type stars and 
including the dense cluster R136.
}
\item{A younger, likely triggered, `Orion Phase' to the north and west of R136.}
\item{A `Sco OB1 Phase' of early-type supergiants throughout the central field.}
\item{An older `h \& $\chi$ Persei Phase' in Hodge~301, containing cooler, more evolved, supergiants, 
to the northwest of the centre.}
\item{A separate `Sco OB1 Phase' surrounding the luminous blue variable R143.}
\end{itemize}
\smallskip
With its rich stellar populations, 30~Dor is the ideal laboratory in
which to investigate a number of important outstanding questions
regarding the physics, evolution, binary fraction, and chemical
enrichment of the most massive stars.  Building on the successes of
the VLT-FLAMES Survey of Massive Stars \cite{e05}, here we introduce a
new multi-epoch spectral survey of over 1,000 massive stars in the 30~Dor region.

In the broader context, 30~Dor is at the northern end of a large
column of molecular gas which extends south for over 2,000\,pc
\cite{c88,f08}.  N-body models examining the recent edge-on
motion of the LMC through the halo of the Milky Way suggest
significant star formation in the eastern part of the LMC, as
manifested by 30~Dor, due to ram pressure \cite{m09}.  With the
reservoir of gas to the south, the region seems destined to become
an even more spectacular star-formation complex over the next
few million years.

\section{Multiplicity in Massive Stars}
The effects of binarity/multiplicity on the formation and subsequent
evolution of high-mass stars is a vibrant area of research.  Indeed,
one of the key ingredients missing from current theories of both star
formation and cluster evolution is a robust binary fraction of massive
stars, and the distribution of the mass ratios in these systems.
Some motivation in this direction was provided by \cite{zy07}:
\begin{quotation}
`The future of spectroscopic massive binary research lies in the near-IR and
in multi-epoch radial velocity surveys of embedded massive stars'
\end{quotation}
These words were primarily concerned with the earliest stages of star
formation but they coincide with growing interest in multi-epoch
spectroscopic studies in open clusters, aimed at identification and
characterisation of their binary populations (Table~\ref{binaries}).
The most pertinent of these is the study of 50 early-type stars in
30~Dor by \cite{b09}.  From Gemini spectroscopy at seven epochs they
found a binary fraction of $\ge$50\%, noting that the
data were not inconsistent with it being 100\%.  Recent multi-epoch
AO-corrected SINFONI observations found (tentative) evidence for a
short-period companion in only one of the six central WR stars at the
core of R136 \cite{schnurr09}, but it is clear that there is a very
rich binary population in 30~Dor.

\vspace*{-0.075in}
\begin{table}[h]
\begin{center}
\caption{Selected multi-epoch spectroscopic surveys in open clusters.}\label{binaries}
\begin{tabular}{lcl}
\hline
Cluster & Binary fraction & Reference \\
\hline
IC\,1805 & $\ge$0.20 & \cite{db06} \\
NGC\,6231 & $\ge$0.63 & \cite{s08} \\
NGC\,6611 & $\ge$0.44 & \cite{s09} \\
NGC\,2244 & $\ge$0.17 & \cite{mahy09} \\
30\,Dor & $\ge$0.50 & \cite{b09} \\
\hline
\end{tabular}
\end{center}
\end{table}

One of the serendipitous aspects of the FLAMES Survey of Massive Stars
was the large number of spectroscopic binaries discovered
(Table~\ref{fsms}).  The time sampling of the service-mode
observations did a reasonable (but not thorough) job of binary
detection, with lower limits to the binary fraction of $\sim$30\% in
three of the target clusters.  Adopting the same methods as
\cite{s09}, we have calculated the detection probabilities for short,
intermediate and long period binaries for each cluster field.  The
aggregated detection probabilities (for systems with periods of two
days to ten years) are given in the final column of
Table~\ref{fsms}. The similarity in the detection probabilities
suggests that the lower fraction found in NGC\,330 is genuinely
different to the others.  While it is unfair to compare the NGC\,330
observations with the rich, younger cluster fields of NGC\,346 and
N11, NGC\,2004 is its LMC cousin; this difference in the binary
fraction is intriguing and the subject of ongoing work.

\vspace*{-0.075in}
\begin{table}[h]
\begin{center}
\caption{Spectroscopic binaries from Evans et al. (2006).}\label{fsms}
\begin{tabular}{lccccc}
\hline
Cluster & Galaxy & \#\,O$+$Early B & \#\,Binary & Binary fraction & Detection prob. [2d-10yr]\\
\hline
NGC 346 & SMC & 103 & 27 & $\ge$\,26\% & 0.66 \\
NGC 330 & SMC & 104 & $\phantom{2}$4 & $\ge$$\phantom{2}$\,4\% & 0.71 \\
NGC 2004 & LMC & 105 & 24 & $\ge$\,23\% & 0.64 \\
N11 & LMC & 120 & 43 & $\ge$\,36\% & 0.64 \\
\hline
\end{tabular}
\end{center}
\end{table}

The relationship of the binary fraction with density and the
spatial extent of a cluster is still unclear (e.g. Mahy et al., 2009),
while the binary fraction in OB associations is often similar to clusters, but with
fewer short-period systems (Zinnecker \& Yorke, 2007).

\section{The Tarantula Survey}

The new survey comprises 160\,hrs of VLT-FLAMES spectroscopy in the 30~Dor
region (PI: Evans).  Most of the observations (142\,hrs) have now been
completed, with the remainder scheduled for the coming semester.

One of the prime drivers for this survey was the issue of binarity, shaping
the multi-epoch observational strategy.  It is clear that
identification of binaries, and the mass ratios in those
systems, is an important empirical result for N-body models of star
and cluster formation/evolution.  Moreover, in many clusters,
e.g. NGC\,6231 (Sana et al., 2008), the majority of O-type stars are
members of a binary system.  Thus, to gain a true understanding of the upper 
H-R diagram, the effects of binarity need
to be fully included in theoretical models of stellar evolution 
\cite{selma}.  Although we have focussed on this aspect for this symposium,
the genesis of the survey arose from a much broader range of other
scientific motivations, including:
\begin{itemize}
\item{The role of stellar rotation in the chemical enrichment and
evolution of massive stars. Hunter et al. (2008) have revealed new
challenges for theory in B-type stars; we seek to 
investigate these effects in the more dominant, massive O-type stars.}
\item{Determination of the rotational velocity distribution in 30~Dor.
Are there sufficient high-mass, rapidly-rotating stars to provide a 
channel for long-duration $\gamma$-ray bursts \cite{y06}?}
\item{Armed with precise radial velocities and identification of binaries, 
do we see kinematic evidence of mass segregation and/or infant
mortality in and around R136?}
\item{A more holistic objective of a near-complete census of the closest `proto-starburst', 
with applications in the context of population synthesis methods
and interpretation of spectra of unresolved massive stars clusters at
Mpc distances.}
\end{itemize}

\section{GIRAFFE Observations}

The primary dataset comprises spectroscopy of 1,000
stars using the GIRAFFE spectrograph, which is fed by 132 MEDUSA
fibres available for science (or sky) observations across a
25$^\prime$ field \cite{flames}.  Targets were selected from
unpublished imaging with the Wide-Field Imager (WFI) on the ESO/MPG
2.2-m telescope, and from Brian Skiff's reworking
of the \cite{selman} photometric catalogue in the central
90$^{\prime\prime}$.  To obtain a representative sample of the upper
part of the HR diagram, including evolved luminous stars, no colour
cut was applied to potential targets but a faint cut-off ($V\,<\,$17)
was enforced to ensure sufficient signal-to-noise for each star.

Nine MEDUSA configurations were observed, each of which was
observed at three wavelength settings (see Table~\ref{data}).  This
yields full coverage of the classical blue-optical region used for
spectroscopic classification and analysis, combined with higher
resolution spectroscopy of the H$\alpha$ region to enable determination of
the stellar wind intensity.  From inspection of initial reductions,
the minimum signal-to-noise in the stacked spectra for the faintest stars is $\sim$50,
i.e. the spectra are suitable for quantitative analysis as well as
radial velocity monitoring.

\begin{table}[h]
\begin{center}
\caption{Summary of FLAMES-GIRAFFE observations.}\label{data}
\begin{tabular}{cccc}
\hline
GIRAFFE setting & $\lambda$-coverage (\AA) & $R$  & Exposures \\
\hline
LR02 & 3980-4535 & 6,500 & 6$\times$(2$\times$1815s) \\
LR03 & 4505-5050 & 7,500 & 3$\times$(2$\times$1815s) \\
HR15N & 6470-6790 & 17,000 & 2$\times$(2$\times$2265s) \\
\hline
\end{tabular}
\end{center}
\end{table}

\begin{figure}[t]
\begin{center}
 \includegraphics[width=5in]{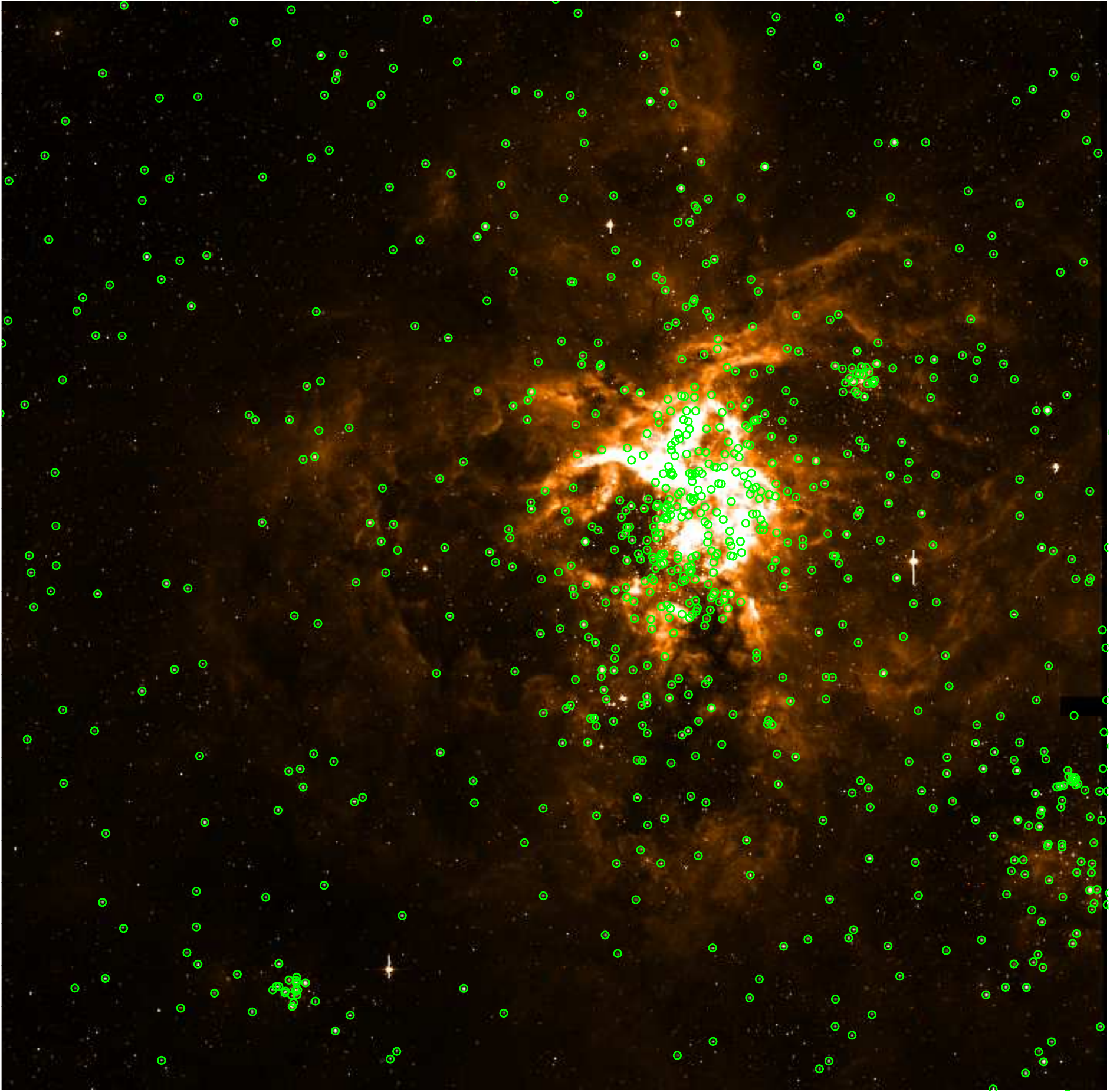} \\ 
\caption{14$^\prime\,\times$14$^\prime$ V-band WFI image showing the
FLAMES-GIRAFFE targets in and around 30~Dor (north to the top, east to the left).}\label{fig1}
\end{center}
\end{figure}

The distribution of the majority of the MEDUSA targets is shown in
Figure~\ref{fig1}.  The survey samples the full extent of 30~Dor and
outwards into the `field' population and other nearby OB associations
to fully exploit the FLAMES field-of-view and spare fibres, thus
bolstering the observational sample.

\subsection{Preliminary Classification}

Work is now progressing in earnest with the final science reductions
of the GIRAFFE spectra.  To characterise the spectral content of the
survey, we extracted the reduced spectra from one pair of LR02
observations for each MEDUSA configuration.  In advance of the
full reductions, it was not possible to classify $\sim$150 stars from just one
observation (although most are likely B-type stars) but from visual
inspection of the spectra the sample contains:
\begin{itemize}
\item{In excess of 300 O-type stars and $\sim$20 Wolf-Rayet/`slash' stars.
This is a hugely significant improvement in terms of sampling the
upper HR diagram, e.g., compared to the analysis of 28 O-type stars
in the LMC by \cite{m07}.  Each star will be studied for binary companions, 
and then analysed to obtain physical and stellar wind parameters, including the
first large-scale study of nitrogen enrichment in O-type stars.}
\item{Over 400 B-type spectra, which will be used to establish the
baseline chemical abundances in 30~Dor and, with such a large sample
in one field, will be used to revisit the role of rotationally-induced
mixing on surface nitrogen enrichment \cite{h08}.}
\item{$\sim$150 cooler stars with spectral types of A, F and later.  Some
will be foreground objects to be discarded, but the majority will be
evolved, luminous stars which will be used to investigate the
short lifetimes of these evolutionary phases via
population synthesis models.}
\end{itemize}

\subsection{Binary Detection Probabilities}

\begin{figure}[h]
\begin{center}
 \includegraphics[width=4in]{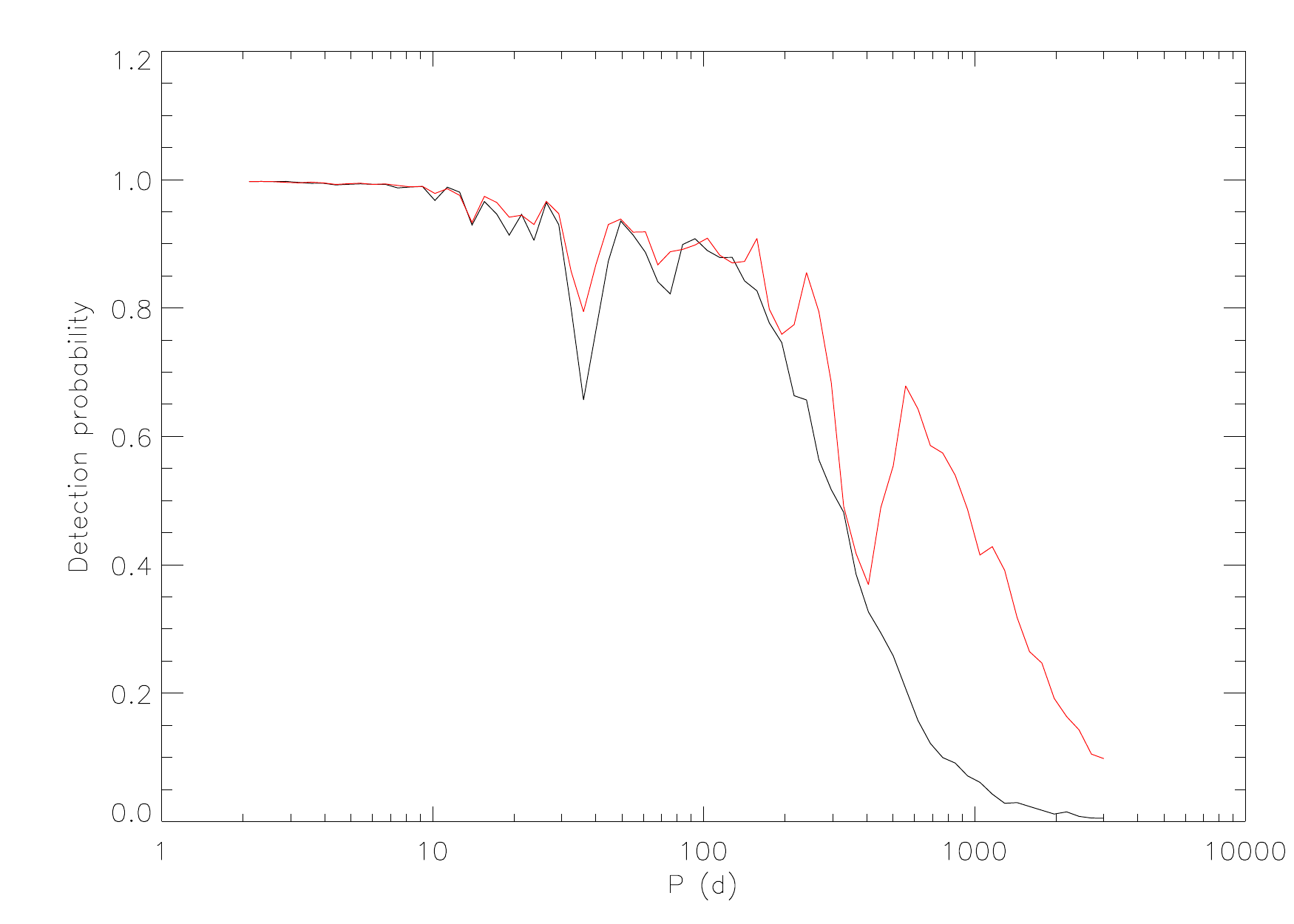} \\ 
 \caption{Detection probability of binary companions for one of the
MEDUSA configurations. The black line shows the detection results for
the five epochs already executed; the red (grey) line illustrates
the increased probability obtained from a sixth epoch, scheduled for
observation in the coming semester, i.e. separated by approximately one year
from the other epochs.}\label{fig2}
\end{center}
\end{figure}

Using the methods from \cite{s09} we have calculated the
detection probabilities for binaries from the actual time sampling of
the nine observed MEDUSA configurations.  In these calculations we
assume a $\Delta v_{\rm r}$ threshold of 20 km$~{\rm s}^{-1}$, requiring
a radial velocity precision of $\sim$5 km$~{\rm s}^{-1}$ (which should be
achievable at the resolving power of the new spectroscopy for all but the 
fastest rotating stars).  The detection probabilities, as a function of orbital
period, for the first MEDUSA configuration are shown by the black line in 
Figure~\ref{fig2}.  We are relatively complete up to periods of a few 10s of
days, with a steep fall-off beyond 100 days.  The inclusion of one additional
epoch in the coming observing season significantly helps with the detection
of both intermediate and long period binaries (red/grey line).
By quantifying our detection probabilities using such simulations, we will be able
to put firm limits on the observed binary fraction.

\vspace{-0.2in}
\section{Supplementary Data}

\subsection{ARGUS \& UVES Observations of R136}
R136 is too dense for effective use of the MEDUSA fibres, so a
15$^{\prime\prime}$ exclusion radius around the core was employed in
the fibre allocations.  To investigate the dynamics and binarity of
stars in and around R136, as part of the Large Programme we have
observed five pointings with the ARGUS integral field unit (IFU) which
delivers a 12$^{\prime\prime}\,\times\,$7$^{\prime\prime}$
field-of-view.  Each pointing has been observed with the LR02 GIRAFFE
setting (which delivers a resolving power of $\sim$10,000 in the IFU
mode) at five epochs.

In parallel to the ARGUS observations, we used the fibre-feed to the
red arm of UVES to observe 25 stars that were not included in the
MEDUSA configurations.  The $\lambda$5200 standard set-up was used,
delivering spectral coverage of $\sim\lambda\lambda$4200-6200 at
$R\,=\,$47,000.

\subsection{VLT-SINFONI K-band Spectroscopy}

The majority of the known WR and extreme O-type emission-line stars in 30~Dor
are in the central regions.  Near-IR IFU observations with SINFONI
(12hrs; PI: Gr\"{a}fener) will be used to obtain K-band spectroscopy
of the central arcminute around R136.  The stellar wind lines in the K-band, 
principally from Brackett~$\gamma$ and He~{\small II}, are more sensitive
than the optical lines at low mass-loss rates, enabling a more precise determination
of the physical parameters of the most extreme stars.

\subsection{Faulkes Photometric Follow-up}
In the longer term, the spectroscopy in the central $\sim$10
arcminutes will be supplemented with multi-band photometric monitoring
from the Faulkes Telescope South, as part of their schools education
programme.  Faulkes has a
4{\mbox{\ensuremath{.\!\!^\prime}}}7\,$\times$\,4{\mbox{\ensuremath{.\!\!^\prime}}}7
field-of-view, so the main body of 30~Dor will be mapped with several
pointings, delivering multi-epoch photometry that will, for example,
assist with the analysis of identified binary systems.

\section{Summary}
We have an exceptional and unique data resource available to us to investigate
the massive-star population in 30\,Dor, now the hard work begins!

\bigskip
\noindent{\bf Acknowledgements:} Based on observations from ESO programme 182.D-0222.
We are indebted to Brian Skiff for his careful reworking of the Selman et al. astrometry.

\end{document}